\documentstyle[11pt]{article}
\begin{document}

%%%%%%%%%%%%%%%%%%%
\def\com#1,#2,#3#4{          {\it Comm. Math. Phys.\/ }{\bf #1} (19#3#4) #2}
\def\np#1,#2,#3#4{           {\it Nucl. Phys.\/ }{\bf B#1} (19#3#4) #2}
\def\npps#1,#2,#3#4{         {\it Nucl. Phys. B (Proc. Suppl.)\/ }{\bf B#1}
                             (19#3#4) #2}
\def\pl#1,#2,#3#4{           {\it Phys. Lett.\/ }{\bf #1B} (19#3#4) #2}
\def\pla#1,#2,#3#4{          {\it Phys. Lett.\/ }{\bf #1A} (19#3#4) #2}
\def\pr#1,#2,#3#4{           {\it Phys. Rev.\/ }{\bf D#1} (19#3#4) #2}
\def\prep#1,#2,#3#4{         {\it Phys. Rep.\/ }{\bf #1} (19#3#4) #2}
\def\prl#1,#2,#3#4{          {\it Phys. Rev. Lett.\/ }{\bf #1} (19#3#4) #2}
\def\pro#1,#2,#3#4{          {\it Prog. Theor. Phys.\/ }{\bf #1} (19#3#4) #2}
\def\rmp#1,#2,#3#4{          {\it Rev. Mod. Phys.\/ }{\bf #1} (19#3#4) #2}
\def\sp#1,#2,#3#4{           {\it Sov. Phys.-Usp.\/ }{\bf #1} (19#3#4) #2}
%%%%%%%%%%%%%%%%%%%%%%%%

\def\thefootnote{\fnsymbol{footnote}}
\def\BK#1#2{{\left[{#1\atop#2}\right]}}
\def\pol{{1 \over 2}}
\def\cns#1#2{\chi_{#1,#2}^{\rm NS}}
\def\chr#1#2{\chi_{#1,#2}^{\rm R}}
\def\DER#1{q^{{1 \over 4} l C_{#1} l}}
\def\DE#1#2{q^{{1 \over 4} l C_{#1} l-\pol #2}}
\def\PROG{\prod_{a=1,3,..k-1} \BK{\pol ( l I_{k-1}+u)_{a}}{l_{a}}}
\def\f{\cns{r}{s}=\prod_{n=1}^{\infty} {{(1+q^{n-\pol})} \over {(1-q^{n})}}
\sum_{j=-\infty}^{\infty} q^{\gamma^{k}_{r,s} (j)}-q^{\beta^{k}_{r,s} (j)}}
\def\F#1#2#3#4{\Phi_{#1}(#2|#3,#4)}
\def\ff{\chr{r}{s}=\prod_{n=1}^{\infty} {{(1+q^{n})} \over {(1-q^{n})}}
\sum_{j=-\infty}^{\infty} q^{\gamma^{k}_{r,s} (j)}-q^{\beta^{k}_{r,s} (j)}}
\def\ste{{{(2 k (k+2)j+r (k+2)+k s)^2-(r (k+2)-k s)^2}\over {4 k (k+2)}}}
\def\st{{{(2 k (k+2)j-r (k+2)+k s)^2-(r (k+2)-k s)^2} \over {4 k (k+2)}}}
\def\mod{\rm mod}
\def\Fe#1#2#3#4{F_{#1}(#2|#3,#4)}
%%%%%%%%%%%%%%%%%%%%%%%%%%%%%%%

\begin{center}
\hfill    WIS--95/5/Feb.--PH\\
\hfill       hep-th/9502118
\vskip 1 cm

{\large \bf  Fermionic Sum Representations for the Virasoro Characters of the
Unitary Superconformal Minimal Models}

\vskip 1 cm

Ernest Baver and Doron Gepner

\vskip 1 cm

{\em Department of Particle Physics\\
The Weizmann Institute\\
Rehovot 76100\\
ISRAEL}

\end{center}

\vskip 1 cm

\begin{abstract}
We present fermionic sum representation for the general Virasoro character of
the unitary minimal superconformal series ($N=1$). Example of the corresponding
``finitizated" identities relating corner transfer matrix polynomials with
fermionic companions is considered. These identities in the thermodynamic limit
lead to the generalized Rogers-Ramanujan identities.

\end{abstract}

\newpage

\section{Introduction}

One of the puzzling features of the two dimensional systems is the appearance
of
 the characters of the fixed point conformal field theories in the expressions
for the local state probabilities of the lattice models. Corner transfer matrix
method \cite{bax} enables to extract information about lattice models at and
away
 of criticality. It turns out that the logarithm of the corner transfer matrix
is proportional to the Virasoro generator $L_0$, {\it i.e.} their spectra have
the same degeneracies and spacing. This is rather striking result, since the
Virasoro algebra is connected with the conformal symmetry, which operates only
at the critical point, in the continuum limit.

 Another closely related mysterious result is the appearance of generalized
Rogers-Ramanujan identities (GRR) for the characters of rational conformal
field
theory.  Recently numerous generalized Rogers-Ramanujan identities [2--7] were
found/conjectured for
 the Virasoro characters. These identities relate standard ``bosonic"
representation with
 the fermionic-like sums of the form:

$$\sum_{{l \hskip 0.3cm {\bf restrictions}}} q^{{1 \over 4} l C l-\pol A l}
\hskip 0.3cm {\prod_{a} \BK{\pol ( l (1-C)+u)_{a}}{l_{a}}}, \eqno(1.1)$$where
some of the components of the vector $u$ may be infinite, $C$ is symmetric
matrix, and $\BK{n}{m}$ are Gaussian polynomials defined by

$$ \BK{n}{m}={{(q)_{n}} \over {(q)_{m} (q)_{n-m}}}, \eqno(1.2)$$

$$ (q)_{n}=\prod_{j=1}^{n} (1-q^{j}), \eqno(1.3)$$for integers $n \ge m \ge 0$
and $\BK{n}{m}=0$ otherwise.  The beautiful formula for the general Virasoro
character of the minimal unitary models was conjectured in \cite{mccoy}. The
natural question which may be asked is: whether a generalization possible for
other models?

In this work we generalize the result obtained in \cite{mccoy,gepner}. We
conjecture the general formula for the Virasoro characters of the unitary
superconformal minimal models generated by the coset construction $SU(2)_{k-2}
\times SU(2)_{2} / SU(2)_{k}$. At the last section we will make few steps
towards the proof of the
conjectured identities, namely we will formulate the stronger conjecture for
the
 ``finitizated" polynomials, which in appropriate limit lead to GRR.

\newpage

\section{General Conjecture}

Standard ``bosonic" representation for the Virasoro characters of the unitary
superconformal minimal models, generated by the coset construction $SU(2)_{k-2}
\times SU(2)_{2} / SU(2)_{k}$, are given by \cite{GKO}:
$$\f \hskip 1cm r-s=0 \hskip 0.1cm {\mod} \hskip 0.1cm 2, \eqno(2.1)$$
$$\ff \hskip 1cm r-s=1 \hskip 0.1cm \mod \hskip 0.1cm 2, \eqno(2.2)$$where

$$\gamma^{k}_{r,s} (j)=\st, \eqno(2.3)$$
$$\beta^{k}_{r,s} (j)=\ste,  \eqno(2.4)$$
$$1 \le r \le k-1, \hskip 0.8cm 1 \le s \le k+1. \eqno(2.5)$$

Let us introduce the following notations:

$$S_{k-1}\BK{Q}{A}(u \mid q)=\sum_{{{l_1 \in Z_{\ge 0}} \atop
l_2,..,l_{k-1}+Q}}
{\DE{k-1}{A l} \over {(q)_{l_2}}}\PROG, \eqno(2.6)$$where $l_2,..,l_{k-1} \in
2
Z_{\ge 0}$, \hskip 0.2cm $Q \in (Z_{2})^{k-2}$ denotes restrictions on
summation
variables $l_2,..,l_{k-1}$,
 $C_{k-1}$ is the cartan matrix of $A_{k-1}$ and $I_{k-1}$ is the incidence
matrix:

$$I_{k-1}=2-C_{k-1},\eqno(2.7)$$

$$(I_{k-1})_{a,b}=\delta_{a,b+1}+\delta_{a,b-1}.\eqno(2.8)$$

Let also $e_{a}$ be unit vector in the $a$ direction {\it i.e.}
$(e_{a})_{b}=\delta_{a,b}$, and set $e_a=0$ for $a\notin \{1,2,...,k-1\}$.

\newpage

It was found in ref. \cite{mccoy} that the identity character of the coset
models $SU(2)_{K} \times SU(2)_{M} / SU(2)_{K+M}$ may be represented in the
form
(1.1) with $A=0$, $C=C_{K+M-1}$, and $u_{M}=\infty$, all other $u_{a}=0$, for
$a=1,...,K+M-1$.

We conjecture\footnote{The expression for the identity character $\chi_{1,1}$
conjectured in ref.\cite{mccoy} is recovered from Eqs.(2.9--2.12).} that
Virasoro characters of the superconformal minimal models may be represented in
the form (2.6) with different characteristics $u$, $Q$ and $A$:

$$\chi_{r,s}={1 \over {(1+\epsilon^{r-s})}} q^{- {1 \over 8 }
(s-r-\epsilon^{r-s})(s-r-2+\epsilon^{r-s})}
S_{k-1}\BK{Q_{r,s}}{A_{r,s}}(u_{r,s}
\mid q), \eqno(2.9)$$

$$A_{r,s}=e_{s-1}, \eqno(2.10)$$

$$u_{r,s}=e_{s-1}+e_{k+1-r}+\epsilon^{r-s} e_1, \eqno(2.11)$$

$$Q_{r,s}=(r-1) \rho +(e_{s-2}+e_{s-4}+...)+(e_{k+2-r}+e_{k+4-r}+...),
\eqno(2.12)$$where $\rho=e_1+...+e_{k-1}$, and

$$\epsilon^{r-s}= {{0 \hskip 0.2cm r-s=0 \hskip 0.1cm {\mod} \hskip 0.1cm 2}
\atop {1 \hskip 0.2cm r-s=1  \hskip 0.1cm {\mod} \hskip 0.1cm 2}}.
\eqno(2.13)$$Here the addition of components in $Q_{r,s}$ is done modulo $2$,
and afterward  the projection on the plane perpendicular to $e_1$ is taken,
{\it
i.e.} the summation variable $l_1$ always takes positive integer values. We
checked this conjecture by computer for low values of $k$ to high order in $q$.

Note that the symmetry of the conformal grid $\chi_{r,s}=\chi_{k-r, k+2-s}$
implies new
identities relating sums of the form $(2.6)$ with different characteristics.

In closing this section we would like to stress the remarkable similarity
between the formula (2.9) and the formula conjectured for the general Virasoro
character of the minimal series \cite{mccoy}.

\newpage
\section{Lattice Models and ``Finitization" of GRR}
The classical way to prove identities (2.9) is to consider first stronger
statement, namely the equality of finite polynomials which after taking
infinite
limit will lead to the generalized Rogers-Ramanujan identities. The
``finitizated" version of GRR identities is known for the appropriate
characters
of the ordinary minimal models. It was conjectured \cite{meltzer}, and in some
cases proved \cite{meltzer,berkovich,warn} that the expressions for the local
state probabilities (before thermodynamic limit is taken) of the
Andrews-Baxter-Forrester models coincide with appropriate finitizated fermionic
expressions of the form (1.1). Below we will give the example showing that
similar finitizated identities exist for the Virasoro characters of the minimal
superconformal series.

 Consider lattice models which in the critical region correspond to the coset
rational conformal field theory $SU(2)_{k-2} \times SU(2)_{2} / SU(2)_{k}$
\cite{kyoto}. The state variables $l$ in these models can take the values
$l=1,2,...,k+1$. Two state variables $l_1$ and $l_2$ siting on the same bond
must obey admissibility condition:

$$l_1 \sim l_2 <=>  {l_1-l_2=-2,0,2 \atop {l_1+l_2=4,6,..,2 k }}.
\eqno(3.1)$$The ground states are labeled by a pair of states, $b$ and $c$,
such
that $b \sim c$. The local state probability is defined as the probability to
find the state $a$ at the $l_{0,0}$ site, for the $(b,c)$ ground state,

$$P(a|b,c)=\langle \delta(a,l_{0,0}) \rangle.\eqno(3.2)$$In the regime III
local
state probability is given in terms of one dimensional configuration sum
\cite{kyoto},

$$\Phi(a|b,c)=\lim_{L\rightarrow \infty} \F{L}{a}{b}{c},\eqno(3.3)$$

$$\F{L}{a}{b}{c}=\sum_{l_{i}} q^{\sum_{j=1}^{L} {j \over 4}
|l_{j+2}-l_{j}|},\eqno(3.4)$$where $l_1=a$, $l_{n+1}=b$, $l_{n+2}=c$, and the
sum goes
 over all admissible sequences $l_1 \sim l_2 \sim ... \sim l_{n+2}$. Here $q$
is
related to the temperature-like parameter of the lattice model.

 It was shown in ref. \cite{kyoto} that the configuration sum $\Phi(a|b,c)$ is
given in terms of branching functions $c_{r s a}$ of this coset rational
conformal field theory $SU(2)_{k-2} \times SU(2)_{2} / SU(2)_{k}$,

$$\Phi(a|b,c)=q^{\nu} c_{r s a},\eqno(3.5)$$
$$r= {1 \over 2 } (b+c-2), \hskip 0.5cm s= {1 \over 2 }
(b-c+4).\eqno(3.6)$$Configuration
 sum $\F{L}{a}{b}{c}$ obeys the following recursion relations:

$$\F{L}{a}{b}{c}=\sum_{d \sim b} q^{{L \over 4} |d-c|} \F{L-1}{a}{d}{b},
\eqno(3.7)$$where the sum is taken over all $d$ admissible with $b$. These
relations together with boundary condition $\F{0}{a}{b}{c}=\delta_{a,b}$
uniquely determines configuration sum $\F{L}{a}{b}{c}$.

As an example we will consider the case $k=4$. Let us introduce the following
notations:

$$\Fe{L}{1}{1}{3}=\sum_{l_1 \atop l_2,l_3 even} q^{{1 \over 4} l C_3 l}
\BK{\pol
l_2}{l_1} \BK{\pol l_2}{l_3} \BK{\pol (l_1+l_3+L)}{l_2},\eqno(3.8)$$

$$\Fe{L}{1}{3}{1}=\sum_{l_1 \atop l_2,l_3 even} q^{{1 \over 4} l C_3 l}
\BK{\pol
l_2}{l_1} \BK{\pol l_2}{l_3} \BK{\pol (l_1+l_3+L+1)}{l_2},\eqno(3.9)$$

$$\Fe{L}{1}{3}{5}=\sum_{l_1 \atop l_2 even, l_3 odd} q^{{1 \over 4} l C_3 l}
\BK{\pol l_2}{l_1} \BK{\pol l_2}{l_3} \BK{\pol
(l_1+l_3+L+1)}{l_2},\eqno(3.10)$$

$$\Fe{L}{1}{5}{3}=\sum_{l_1 \atop l_2 even, l_3 odd} q^{{1 \over 4} l C_3 l}
\BK{\pol l_2}{l_1} \BK{\pol l_2}{l_3} \BK{\pol
(l_1+l_3+L)}{l_2}.\eqno(3.11)$$Now we may formulate the following conjecture:

$$\Fe{L}{1}{1}{3}=\F{L}{1}{1}{3},\hskip 0.8cm
\Fe{L}{1}{3}{1}=\F{L}{1}{3}{1},\eqno(3.12)$$
$$\Fe{L}{1}{3}{5}=\F{L}{1}{3}{5},\hskip 0.8cm
\Fe{L}{1}{5}{3}=\F{L}{1}{5}{3},\eqno(3.13)$$where $\F{L}{a}{b}{c}$ are defined
from Eq. (3.4). In order to prove this statement we
 have to show that $\Fe{L}{a}{b}{c}$ obey recursion relation $(3.7)$ together
with the initial condition.

It is easy to show that $\Fe{L}{a}{b}{c}$ defined above satisfy the same
initial
condition as $\F{L}{a}{b}{c}$, namely:

$$\Fe{0}{1}{1}{3}=\F{0}{1}{1}{3}=1, \eqno(3.14)$$
$$\Fe{0}{1}{3}{1}=\F{0}{1}{3}{1}=0, \eqno(3.15)$$
$$\Fe{0}{1}{5}{3}=\F{0}{1}{5}{3}=0, \eqno(3.16)$$
$$\Fe{0}{1}{3}{5}=\F{0}{1}{3}{5}=0. \eqno(3.16)$$Let us rewrite the equations
$(3.7)$ relevant for our case in explicit form:

$$\F{L}{1}{3}{3}=\F{L-1}{1}{3}{3}+\F{L-1}{1}{1}{3} q^{L \over
2}+\F{L-1}{1}{5}{3} q^{L \over 2},\eqno(3.17)$$

$$\F{L}{1}{3}{1}=\F{L-1}{1}{1}{3}+\F{L-1}{1}{3}{3} q^{L \over
2}+\F{L-1}{1}{5}{3} q^{L},\eqno(3.18)$$

$$\F{L}{1}{3}{5}=\F{L-1}{1}{5}{3}+\F{L-1}{1}{3}{3} q^{L \over
2}+\F{L-1}{1}{1}{3} q^{L},\eqno(3.19)$$

$$\F{L}{1}{1}{3}=\F{L-1}{1}{3}{1},\eqno(3.20)$$

$$\F{L}{1}{5}{3}=\F{L-1}{1}{3}{5}.\eqno(3.21)$$It is obvious that
$\Fe{L}{a}{b}{c}$ defined above Eqs. (3.8-3.11) satisfy equations (3.20) and
(3.21). It was checked for numerous values of parameter $L$ that finitizated
fermionic expressions $\Fe{L}{a}{b}{c}$ Eqs. (3.8-3.11) do coincide with
appropriate $\F{L}{a}{b}{c}$.

Taking limit $L \rightarrow \infty$ in equations (3.12-3.13) we obtain GRR
identities for the branching functions of the coset $SU(2)_{k-2} \times
SU(2)_{2} / SU(2)_{k}$. For example:

$$c_{111}=\lim_{L\rightarrow \infty} \Fe{L}{1}{1}{3}=\sum_{l_1,l_2,l_3 even}
{q^{ {1\over 4} l C_3 l} \over (q)_{l_2} } \BK{\pol l_2}{l_1} \BK{\pol
l_2}{l_3}, \eqno(3.22)$$

$$c_{131}=\lim_{L\rightarrow \infty} \Fe{L}{1}{3}{1}=\sum_{l_1 odd, \atop
{l_2,l_3 even}} {q^{ {1\over 4} l C_3 l} \over (q)_{l_2} } \BK{\pol l_2}{l_1}
\BK{\pol l_2}{l_3}, \eqno(3.23)$$

where the limit is taken for $L$ even and we used
$$\lim_{L\rightarrow \infty} \BK{L+a}{n}={1 \over (q)_n}.
\eqno(3.24)$$Similarly
we have:

$$c_{311}=\lim_{L\rightarrow \infty} \Fe{L}{1}{3}{5}=\sum_{{l_1,l_2 even} \atop
l_3 odd} {q^{ {1\over 4} l C_3 l} \over (q)_{l_2} } \BK{\pol l_2}{l_1} \BK{\pol
l_2}{l_3}, \eqno(3.25)$$

$$c_{331}=\lim_{L\rightarrow \infty} \Fe{L}{1}{5}{3}=\sum_{{l_2 even} \atop
{l_1,l_3 odd}} {q^{ {1\over 4} l C_3 l} \over (q)_{l_2} } \BK{\pol l_2}{l_1}
\BK{\pol l_2}{l_3}. \eqno(3.26)$$From these identities immediately follow
identities for the Virasoro characters:

$$\cns{1}{1}=c_{111}+c_{131}=\sum_{{l_1 } \atop {l_2,l_3 even}} {q^{ {1\over 4}
l C_3 l} \over (q)_{l_2} } \BK{\pol l_2}{l_1} \BK{\pol l_2}{l_3},\eqno(3.27)$$
$$\cns{1}{5}=c_{311}+c_{331}=\sum_{{l_1 } \atop {l_2 even,l_3 odd}} {q^{
{1\over
4} l C_3 l} \over (q)_{l_2} } \BK{\pol l_2}{l_1} \BK{\pol
l_2}{l_3}.\eqno(3.28)$$

\vskip 0.8cm

We believe that results presented here may be generalized to other models. In
particular
 similarity between the fermionic sum representations for minimal and
superconformal minimal models inspires to look for GRR related to the branching
functions/characters of the more general rational coset field theory $SU(2)_{K}
\times SU(2)_{M} / SU(2)_{M+K}$.

\end{document}